# Subwavelength Imaging using a Solid-Immersion Diffractive Optical Processor

*Jingtian Hu,[1,2,3,4#] Kun Liao,[2,3,4,5#] Niyazi Ulas Dinc,[6] Carlo Gigli,[6] Bijie Bai,[2,3,4] Tianyi Gan,[2] Xurong Li,[2] Hanlong Chen,[2,3,4] Xilin Yang,[2] Yuhang Li,[2] Çağatay Işıl,[2] Md Sadman Sakib Rahman,[2] Jingxi Li,[2] Xiaoyong Hu,[5] Mona Jarrahi,[2] Demetri Psaltis,[6] and Aydogan Ozcan\*[2,3,4,7]*

[1]Guangdong Key Laboratory of Semiconductor Optoelectronic Materials and Intelligent Photonic Systems, Harbin Institute of Technology (Shenzhen), Shenzhen, 518055, China

[2]Electrical and Computer Engineering Department, University of California, Los Angeles, California 90095, United States.

[3]Bioengineering Department, University of California, Los Angeles, California 90095, United States.

[4]California NanoSystems Institute (CNSI), University of California, Los Angeles, California 90095, United States.

[5]Department of Physics, Peking University, Beijing, 100871, China

[6]School of Engineering, École Polytechnique Fédérale de Lausanne (EPFL), Switzerland

[7]David Geffen School of Medicine, University of California, Los Angeles, California 90095, United States

\* ozcan@ucla.edu

[#] equal contribution

**Abstract**

Phase imaging is widely used in biomedical imaging, sensing, and material characterization, among other fields. However, direct imaging of phase objects with subwavelength resolution remains a challenge. Here, we demonstrate subwavelength imaging of phase and amplitude objects based on all-optical diffractive encoding and decoding. To resolve subwavelength features of an object, the diffractive imager uses a thin, high-index solid-immersion layer to transmit high-frequency information of the object to a spatially-optimized diffractive encoder, which converts/encodes high-frequency information of the input into low-frequency spatial modes for transmission through air. The subsequent diffractive decoder layers (in air) are jointly designed with the encoder using deep-learning-based optimization, and communicate with the encoder layer to create magnified images of input objects at its output, revealing subwavelength features that would otherwise be washed away due to diffraction limit. We demonstrate that this all-optical collaboration between a diffractive solid-immersion encoder and the following decoder layers in air can resolve subwavelength phase and amplitude features of input objects in a highly compact design. To experimentally demonstrate its proof-of-concept, we used terahertz radiation and developed a fabrication method for creating monolithic multi-layer diffractive processors. Through these monolithically fabricated diffractive encoder-decoder pairs, we demonstrated phase-to-intensity ($P \rightarrow I$) transformations and all-optically reconstructed subwavelength phase features of input objects (with linewidths of ~$\lambda/3.4$, where $\lambda$ is the illumination wavelength) by directly transforming them into magnified intensity features at the output. This solid-immersion-based diffractive imager, with its compact and cost-effective design, can find wide-ranging applications in bioimaging, endoscopy, sensing and materials characterization.

**Keywords:** *diffractive processors, solid immersion imaging, phase-to-intensity transformations*

## 1. Introduction

The ability to extract the phase information from an input optical wavefront with high spatial resolution is critical for various applications ranging from holographic displays [1] to bioimaging [2-5] and materials characterization [6]. Phase imaging of weakly scattering objects such as cells and tissue, for example, plays a key role in fundamental studies of biological systems [7] as well as medical applications, including disease diagnosis [4,8]. Motivated by these applications, various phase imaging methods, such as phase contrast imaging [9] and differential interference contrast (DIC) microscopy [10], have been developed. Phase imaging techniques through scattering media have also been demonstrated by solving the inverse-scattering problem *via* transmission-matrix-based approaches [11-14]. However, the digital reconstruction algorithms behind these phase imaging techniques are, in general, computationally expensive and take relatively long, even with graphics processing unit-accelerated computing [11]. In addition, these methods typically cannot capture quantitative phase information of specimens [15]. To address this limitation, quantitative phase imaging (QPI) techniques have been developed to provide accurate phase information based on a variety of phase-retrieval techniques, including, e.g., digital holography [16-21] and iterative multi-frame reconstruction methods such as ptychography [22,23]. These techniques, however, need intensive post-processing using a computer, which makes the imaging process time-consuming. Moreover, resolving subwavelength phase features is, in general, a challenge for these approaches due to the limited numerical aperture (NA) of such interferometric systems.

To improve the NA of an imaging system, the solid-immersion principle [24-27] can be used to achieve single-shot imaging with subwavelength resolution by placing a high-index material between the object/specimen and the objective lens [28-30]. However, despite their advanced set-up, solid-immersion microscopy systems have not yet realized subwavelength phase imaging.

Oil/water immersion-based microscopy systems [31-33] that use high-index liquids around the object/specimen were also demonstrated. However, these techniques either exhibit limited phase-imaging resolution despite using structured illumination [34,35] or require off-axis illumination with relatively intensive digital post-processing using a computer, which makes the imaging process time-consuming [36].

Here, we report a compact, solid-immersion imaging framework to achieve subwavelength resolution using all-optical diffractive processors that can realize phase-to-intensity ($P \rightarrow I$) and intensity-to-intensity ($I \rightarrow I$) transformations at the subwavelength scale. The encoder layer of this optical processor is designed to transform/encode the high-frequency information received from the object *via* a high-index medium (refractive index $n > 1$) into lower frequency spatial modes that propagate in air. The subsequent diffractive decoder, which is jointly trained with the encoder surface, processes the encoded spatial information through the air to synthesize at its output plane a magnified image of the input object, revealing subwavelength features that would normally be washed out due to the limited NA in air. When blindly tested with various objects, including subwavelength phase and amplitude structures, this encoder-decoder pair successfully resolved spatial features with a linewidth of $\sim\lambda/2n$, which could not be achieved without the solid-immersion frequency encoder. Notably, the trained subwavelength diffractive imager generalized not only to previously unseen objects from the same distribution as the objects used in training (internal generalization), but also to new types of objects from completely different datasets, demonstrating external generalization capability.

To experimentally demonstrate the feasibility of this subwavelength diffractive imaging platform, we fabricated a multi-layer monolithic design that operates at the terahertz part of the spectrum. We tested this monolithic diffractive encoder-decoder pair with a customized high-resolution terahertz imaging system using a microprobe-based time-domain spectroscopy (TDS) system. Our experimental results confirmed that this 3D-fabricated solid-immersion diffractive imager can resolve phase objects (directly performing $P \rightarrow I$ transformations through the diffractive encoder-decoder pair), revealing subwavelength phase features corresponding to linewidths of ~$\lambda/3.4$ that would normally be lost due to the limited NA in air.

There are several important aspects of this work: the presented solid-immersion diffractive imager has a very compact design that axially spans less than $100\lambda$; this compact design demonstrates $P \rightarrow I$ transformations, performing direct (all-optical) quantitative phase retrieval at the subwavelength scale through the encoding of higher spatial frequencies that travel in a high index dielectric medium. Furthermore, these solid-immersion diffractive processor-based subwavelength imagers can operate at different parts of the electromagnetic spectrum by physically scaling (i.e., expanding or shrinking) the optimized diffractive features of the encoder/decoder surfaces in proportion to the illumination wavelength, $\lambda$, and this is achieved *without* the need for redesigning the diffractive features. We believe that solid-immersion diffractive optical processors, with their subwavelength imaging and high spatial frequency processing capabilities, would provide highly compact and cost-effective solutions for various applications in, e.g., bioimaging, sensing, and material inspection, among many others.

## 2. Results and Discussion

### 2.1 Subwavelength imaging using solid-immersion diffractive optical processors

We first established a numerical model of our solid-immersion subwavelength imager as a diffractive network that is designed *via* a deep-learning-based training process. **Fig. 1** shows the operational principles and the building blocks of our diffractive subwavelength imager based on the solid-immersion principle. As depicted in **Fig. 1a**, the input object with subwavelength features is placed on a high-index dielectric slab positioned in front of the diffractive encoder surface. The function of this diffractive optical encoder is to transform/encode the high-frequency information ($f > 1/\lambda$) of the object into low-frequency representations ($f \leq 1/\lambda$) that can transmit in air, where $f$ represents the spatial frequencies that make up the object information, as depicted in **Fig. 1b**. These encoded optical fields propagate in air and are subsequently processed by a series of all-optical diffractive decoders positioned in air to recover the subwavelength features of the input object with a magnification factor of M (at the output plane). During the design process of this diffractive solid-immersion imager, these *decoder* diffractive layers that pass information through the air were jointly trained with the solid-immersion *encoder* layer, enabling the system to learn an effective encoding-decoding transformation to defeat the bottleneck of the diffraction limit in air. For optimal imaging performance, the axial distance in air between the diffractive encoder and the first decoder layer and between any two consecutive decoder layers were empirically set to ~$12\lambda$; the axial distance between the last diffractive layer and the output plane was empirically set to ~$16\lambda$. Each encoder/decoder layer has 120×120 diffractive neurons/features, each with a lateral size of ~$0.53\lambda$ and a trainable transmission phase coefficient covering 0-$2\pi$ (**Fig. 2a**). The diffractive encoder and decoder layers were jointly optimized using a deep-learning-based training process using datasets composed of custom-designed gratings and EMNIST digits/letters (see the Methods section for details). The optimized phase structures of the encoder-decoder layers for the *I* → *I* and *P* → *I* imaging tasks are presented in **Fig. 2b**. For both designs, the axial thickness of the high-index ($n = 1.72$) immersion material between the object and the all-optical encoder was set to be $1\lambda$ (**Fig. 2a**).

After the training process, we first numerically demonstrated the performance of the solid-immersion-based encoder-decoder pair for subwavelength imaging of various input test objects, including e.g., gratings and EMNIST letters that were never seen before during the training process. **Fig. 2** demonstrates the performance (blind testing results) of two diffractive processors, each with 5 phase-only decoder layers, trained to image amplitude and phase objects with a magnification factor of M = 3, through $I \rightarrow I$ and $P \rightarrow I$ transformations, respectively. Furthermore, we quantitatively evaluated the performance of these solid-immersion diffractive optical processors by comparing their output images with the ground truth by calculating the mean squared error (MSE) and the structural similarity index measure (SSIM) used as image quality metrics. Through these metrics, in **Fig. 2c**, we report the overall imaging performance for both $I \rightarrow I$ and $P \rightarrow I$ tasks using resolution test targets (with various linewidths) that were not used during the training. Limited by the input numerical aperture (NA = $n$ = 1.72), the imaging performance for both the amplitude and phase objects decreased when the linewidths of the test objects were reduced to less than $\lambda/2n$, indicated with the vertical dashed lines in **Fig. 2c**. Interestingly, the $P \rightarrow I$ solid-immersion diffractive imager provided slightly better output MSE and SSIM values compared to the $I \rightarrow I$ diffractive imager, highlighting the capabilities of the diffractive network on phase imaging; this $P \rightarrow I$ diffractive imager directly transforms subwavelength *phase* structures of the input objects into output *intensity* patterns, performing a form of all-optical phase retrieval at the subwavelength level.

To verify the essential role of the diffractive solid-immersion encoder surface in these subwavelength imaging results, we also compared the performances of the presented encoder-decoder systems against decoder-only diffractive imagers, i.e., without the solid-immersion encoder (see **Fig. 2c**, dashed lines). To make this comparison fair, the arrangement of the diffractive decoder layers in air is kept the same (see Supplementary Fig. S1 for details). The dashed lines in **Fig. 2c** that quantify the performance of the diffractive imagers optimized *without* the encoder layer reveal a significant sacrifice in the output imaging performance of the system, clearly emphasizing the importance of the solid-immersion encoder layer for subwavelength imaging.

We also performed an additional test to confirm the critical role of the subsequent decoder layers in our diffractive imager designs by comparing the performance of a solid-immersion encoder layer that is trained without any decoder layers (see Supplementary Fig. S2). This comparison clearly revealed that a deep learning-optimized encoder layer alone, despite the presence of the same solid-immersion layer with $n = 1.72$, failed to perform subwavelength imaging, once again confirming the essential role of the optimized *collaboration* between the diffractive encoder and decoder layers (see Supplementary Fig. S2).

**Fig. 3** reports some additional examples of the blind testing results of our solid-immersion diffractive imager design, performing $\boldsymbol{P \rightarrow I}$ transformations using new resolution test targets and EMNIST handwritten digits/letters that were not used in the training process. The line patterns of the original phase images were resolved up to a linewidth ~$0.25\lambda$ although the contrast of the output image decreased as the linewidth decreased below $\lambda/3.4$ (**Fig. 3a**). We also verified that the same solid-immersion diffractive imager design can successfully reconstruct new letters and digits from the EMNIST test dataset (see **Fig. 3b**). We obtained similar subwavelength imaging results for intensity-encoded input objects by the diffractive encoder-decoder design shown in the upper panel of **Fig. 2b** which was trained for $\boldsymbol{I \rightarrow I}$ imaging task (see Supplementary Fig. S3).

To demonstrate the external generalization capability of our solid-immersion diffractive imagers, we further numerically tested the trained diffractive encoder-decoder system with additional datasets such as the Fashion-MNIST [37] and QuickDraw [38] in performing $\boldsymbol{P \rightarrow I}$ imaging tasks with new types of objects (see **Fig. 4**). Our analyses revealed that, despite the slightly larger background noise and blurring of some of the finer features, the contours of the Fashion-MNIST images were well reconstructed by our all-optical decoder as shown in **Fig. 4a**. The reconstructed images from the QuickDraw test dataset also correctly revealed various subwavelength features of the input objects (see **Fig. 4b**). These blind-testing results demonstrated the external generalization of our diffractive solid-immersion imagers, highlighting their capabilities for general purpose subwavelength imaging. We also validated that our diffractive solid-immersion designs showed the same external generalization capability for $I \rightarrow I$ imaging tasks, covering new types of intensity-encoded objects never seen before, as illustrated in Supplementary Fig. S4.

These analyses reveal that within the input field-of-view defined by the input aperture of the solid-immersion diffractive processor, we can faithfully recover various subwavelength features of the input objects at different locations and orientations as long as they remain in the imaging field-of-view.

**2.2 Impact of the object-to-encoder distance ($d_1$)**

The encoder surface plays a key role in our solid-immersion diffractive imager designs by transforming the high-frequency information of the object within the high-index dielectric material to low-frequency modes that can propagate in air. Since the encoder surface converts the *propagating* modes within the dielectric material into *propagating* modes in air (to be processed by the successive diffractive decoder layers for all-optical image reconstruction), we expect that its utility and function should be, by and large, independent of the axial distance between the object and encoder surface. To confirm this hypothesis, we studied the imaging performance of the diffractive solid-immersion imager with a range of object-to-encoder distances ($d_1$), which is determined by the axial thickness of the dielectric material (see Supplementary Fig. S5). For this, we trained a series of diffractive imagers with $d_1 = 1-16\lambda$ and evaluated their imaging performance with various resolution test targets (see Supplementary Fig. S5a). As expected, the SSIM values calculated for the reconstructions of the resolution test targets with various linewidths covering $w = 0.253-0.333\lambda$ showed no strong dependence on $d_1$ since the optimized encoder surface transforms the traveling waves within the dielectric solid-immersion material (see Supplementary Fig. S5b). However, one can observe a slight improvement in the output image quality for all the resolution test targets as $d_1$ is increased from $1\lambda$ to $\sim 8\lambda$. The main reason for this slight improvement is the better utilization and optimization of the diffractive features and the degrees of freedom at the encoder surface: for a very small $d_1$, the object plane communicates inefficiently with the optimizable diffractive features located at the edges of the encoder surface, which effectively reduces the trainable degrees of freedom at the encoder (see Supplementary Fig. S5c). Further increase of $d_1$ to $\sim 16\lambda$ caused a relative degradation of the output image quality since the NA of the encoder accordingly decreased.

**2.3 Impact of the number ($L$) of diffractive decoder layers**

Previous theoretical analysis and empirical studies showed that deeper diffractive processors can perform an arbitrarily selected complex-valued linear transformation more accurately and exhibit improved generalization capability for various all-optical statistical inference tasks [39,40]. To shed more light on this depth feature of a diffractive optical processor, here we analyze the impact of the number ($L$) of trainable decoder layers of a solid-immersion diffractive imager on its subwavelength imaging performance. **Fig. 5** presents our quantitative performance analysis for the imaging output of solid-immersion diffractive imagers composed of different numbers of decoder layers ($L = 1-6$) optimized for performing $\boldsymbol{P} \rightarrow \boldsymbol{I}$ imaging tasks; all the rest of the design features were kept the same as before. **Fig. 5b** shows the SSIM values calculated for the output images reconstructed from phase-only resolution test targets with linewidths of $w = 0.253–0.333\lambda$ using $L = 1–6$ decoder layers. These results reveal the depth advantages of the diffractive decoder system, providing better resolution and image quality with a larger number of diffractive layers; for example, the sub-wavelength imaging performance of the encoder-decoder pair decreased dramatically when $L < 3$, indicating the importance of the diffractive network's depth for the all-optical decoding of subwavelength information. **Fig. 5c** further reports sample resolution test target images that were all-optically reconstructed by the diffractive decoders composed of different numbers of layers. As shown in Fig. 5c, the single-layer decoder ($L = 1$) achieved poor image reconstruction quality where the phase features were barely recovered for the smaller linewidths ($w = \sim 0.253–0.333\lambda$). With the addition of a second decoder layer ($L = 2$), the sub-wavelength imaging quality improved significantly, but still it showed degradation in the image contrast for finer features, corresponding to linewidths of e.g., $w = \sim 0.267\lambda$. The reconstruction quality continued to improve with the increasing number of decoder diffractive layers ($L > 2$), as summarized in **Fig. 5b**.

We also tested the generalization capability of these solid-immersion diffractive imagers that were optimized with different $L$ using new test images from internal datasets (EMNIST) and external datasets (Fashion-MNIST and QuickDraw). For high-quality reconstruction of phase objects that were sampled from these datasets, $L > 2$ decoder layers were required (see Supplementary Fig. S6), similar to our earlier results reported in **Fig. 5c**, once again highlighting the importance of architectural depth for ensuring the generalization capability of solid-immersion diffractive optical imagers.

**2.4 Solid-immersion diffractive imager designs with different magnification factors (M)**

To demonstrate the versatility of the presented solid-immersion diffractive imager, we also performed a quantitative evaluation of the output imaging quality for $\boldsymbol{P} \rightarrow \boldsymbol{I}$ diffractive imager designs that cover different magnification factors, M = 1.2–5; we used the same architectural design with 1 diffractive encoder layer and $L = 5$ diffractive decoder layers, collectively performing $\boldsymbol{P} \rightarrow \boldsymbol{I}$ imaging tasks (see Supplementary Fig. S7a). As depicted in Supplementary Fig. S7b,c, the SSIM values of the outputs were lower at M = 1.2 since the targeted output images had relatively smaller linewidths for a smaller M value of 1.2, which could not be effectively resolved in air. As expected, the output SSIM values evaluated for all the resolution test targets improved as the magnification factor M increased from 1.2 to 2 since the feature sizes of the output/magnified images increased above the diffraction limit in air $\sim\lambda/2$. However, a further increase of the magnification factor to M = 5 caused a gradual decrease in output SSIM values for all the resolution test targets with $w = \sim 0.253$–$0.333\lambda$ linewidths; this performance degradation is mainly due to the significant area increase at the output field-of-view for M = 5, which caused aberrations because of the spatially-varying effective numerical aperture of the output image plane (see Supplementary Fig. S7c).

Despite some limitations in performance, these analyses confirm that our solid-immersion diffractive processors could successfully resolve linewidths of $\sim\lambda/3.4$ for magnification factors of M = $\sim 1.7$–5, demonstrating the versatility of our diffractive encoder-decoder designs in reconstructing output images with different magnification factors.

**2.5 Trade-off between output diffraction efficiency ($\eta$) and imaging performance**

Another critical metric to evaluate for a solid-immersion diffractive imager is the output power efficiency $\eta$, defined as the ratio of the total power distributed at the detector pixels divided by the total input power. Our forward propagation model assumes that the diffractive encoder/decoder layers are composed of transmissive dielectric materials with negligible optical absorption, which is a fair approximation considering the small axial thickness of our designs (also see Section 2.7). The diffraction efficiency of our solid-immersion imager designs can be optimized and accordingly enhanced by adding a diffractive efficiency-related loss term during the training stage (see the Methods section for details). This additional loss term that is in favor of improved diffraction efficiency, however, creates an imaging performance trade-off. **Fig. 6a** summarizes the SSIM values calculated for $P \rightarrow I$ image reconstructions of various resolution test targets performed by solid-immersion diffractive imagers, composed of 1 encoder and $L = 5$ decoder layers, jointly trained to achieve different output diffraction efficiencies, covering $\eta = 1.9\% - 20.3\%$. Notably, the average diffraction efficiency of the encoder-decoder system was increased by more than 5-fold (from $\eta = 1.9\%$ to $\eta = 10.5\%$) with a negligible compromise in imaging resolution and contrast, highlighting the capability of our solid-immersion diffractive imager in achieving power-efficient sub-wavelength imaging. **Fig. 6b** further shows the output images of various resolution test targets, which confirm the decent sub-wavelength imaging performance of the diffractive design with $\eta = 10.5\%$. As reported in the last column of Fig. 6b, a further increase in the diffraction efficiency to $\eta > 20\%$ caused degradation in the imaging quality.

**2.6 Impact of solid-immersion refractive index ($n$) on the imaging resolution**

To investigate the impact of the refractive index of the dielectric material between the object and the encoder surface, we performed additional numerical testing of phase-encoded resolution test targets with a large range of $n = 1.1-3.0$; the architecture of the solid-immersion diffractive imager remained the same, consisting of 1 encoder and $L = 5$ decoder diffractive layers (see Supplementary Fig. S8). As expected, our analyses revealed a significant improvement in the imaging resolution with higher $n$: initially resolving linewidths of ~$0.333\lambda$ at $n < 1.3$, we were able to resolve a linewidth of ~$0.253\lambda$ at $n = 2.0$, as shown in Supplementary Fig. S8a. However, a further increase in the refractive index to $n = 2.4$ and $n = 3.0$ did not result in better spatial resolution (see Supplementary Fig. S8b). We attribute this bottleneck to the relatively large lateral size of the encoder diffractive features (~$0.53\lambda$), which cannot effectively process all the propagating high-spatial frequencies that are supported at $n = 2.4$ or $n = 3.0$. In fact, the output imaging resolution can be further improved by reducing the lateral feature size of the trainable encoder layer to ~$0.27\lambda$: in this case, using a solid-immersion dielectric material of $n = 2.4$, the encoder-decoder design could resolve linewidths of ~$0.21\lambda$ as illustrated in Supplementary Fig. S9. Therefore, by adopting appropriate design parameters and diffractive feature sizes for the encoder-decoder pair, the imaging resolution of our solid-immersion diffractive imager can be further improved using dielectric materials with even higher refractive index values, making the presented approach a promising technique for super-resolution imaging of deeply subwavelength structures.

## 2.7 Experimental demonstration of subwavelength phase imaging ($P \rightarrow I$)

We experimentally demonstrated the subwavelength imaging capability of the solid-immersion diffractive imager framework using a fabricated encoder-decoder pair designed for $P \rightarrow I$ imaging at 0.4 THz ($\lambda = 0.75$ mm). This proof-of-concept diffractive optical processor was composed of 1 encoder layer and $L = 2$ decoder layers that were jointly trained using a hybrid dataset composed of resolution test targets and MNIST [41] handwritten digits (see the Methods for details). To physically create this diffractive imager, we developed a fabrication method that produced a monolithic design of the multi-layered diffractive encoder-decoder pair *via* a single 3D-printing session, followed by a clean-up of the support materials between the diffractive layers with a combination of mechanical and chemical processes (see the Methods section and Supplementary Fig. S10 for details). This monolithic fabrication method, by and large, eliminated undesirable misalignments in both the axial and lateral directions between the diffractive layers, which could otherwise cause severe degradation of performance [42]. **Fig. 7a** shows the design of the subwavelength imager consisting of a solid-immersion diffractive encoder and a two-layer diffractive decoder in air. Starting from the phase-delay distributions optimized for each encoder/decoder layer, we calculated the corresponding height profiles to be fabricated (for producing the needed phase distributions). After their 3D printing and the subsequent cleaning processes (see Supplementary Fig. S10), we obtained the solid-immersion diffractive imager in a monolithic design where the encoder/decoder layers were well-aligned in all directions (see **Fig. 7b**). A comparison between the optimized phase profiles of the encoder/decoder layers shown in **Fig. 7c** and the photos of 3D-printed layers in **Fig. 7d** reveal the decent quality of our monolithic fabrication process, confirming that the diffractive layers were not damaged by the cleaning process (Supplementary Fig. S10).

To experimentally test our 3D-printed solid-immersion diffractive imager, we built a customized high-resolution terahertz imaging system based on a microprobe (TeraSpike TD-800-X-HRS, Protemics GmbH, Germany) and a TDS system using a plasmonic nanoantenna array-based terahertz source [43]. The photograph and schematics of the experimental set-up are shown in **Fig. 7e-f**. Each sample to be imaged was mounted on a 3-axis electric motor stage to perform the scanning process, while the signal of each scanned point was detected by a stationary terahertz microprobe with a small tip size of ~2 μm. The combination of the microprobe and the plasmonic nanoantenna array source exhibited a high sensitivity with a signal noise ratio (SNR) of ~90 dB at 0.4 THz (see Supplementary Fig. S11), ensuring high-resolution imaging of the output field of view of the diffractive imager.

To experimentally validate the imaging resolution of our subwavelength imager for $P \rightarrow I$ tasks, we first tested the fabricated monolithic diffractive processor using phase-only input objects with periodic line patterns. **Fig. 8** shows the $P \rightarrow I$ experimental imaging results of the reconstructed horizontal and vertical gratings corresponding to phase-encoded lines with linewidths of ~$0.333\lambda$ and ~$0.293\lambda$. Importantly, all the line patterns were reconstructed with good quality, confirming the subwavelength resolution of our solid-immersion diffractive processor that is 3D-printed. These results also experimentally confirm direct phase retrieval of subwavelength features through $P \rightarrow I$ transformations all-optically performed by the diffractive encoder-decoder pair. Some of the deviations observed between the measurement results and the ground truth might be attributed to the experimental errors introduced by 3D fabrication imperfections and potential misalignments, especially in the axial direction of the output/image plane (see Supplementary Figs. S12-S13 for details). Additional experimental results for successful imaging of various phase-encoded resolution test targets with broader linewidths (~$0.4\lambda$ and ~$0.367\lambda$) are also shown in Supplementary Fig. S14, demonstrating the outstanding $P \rightarrow I$ imaging quality of the solid-immersion diffractive processor for features of different sizes.

We also selected some oblique and curved gratings to further demonstrate the performance of the 3D-printed subwavelength diffractive imager for imaging objects with more complex shapes, which are shown in **Fig. 9a**. Although a slight decrease in the imaging contrast was observed, the diffractive system was still able to image and distinguish each line of the test gratings clearly. To further demonstrate the external generalization capability of our $P \rightarrow I$ design, we tested it with phase-encoded objects selected from the EMNIST test dataset, never used during our training stage; we selected handwritten capital letters "U", "C", "L", "A", which were successfully imaged by our 3D-printed solid-immersion diffractive imager as demonstrated in **Fig. 9b**. The measurement results showed high consistency with both the input phase images and the simulation results, further validating the ability of our solid-immersion diffractive imager to perform subwavelength imaging tasks for general objects. One interesting observation that is particularly visible in **Fig. 9b** is that some of the subwavelength features of the objects are better resolved in our experimental results compared to their numerical counterparts (see, e.g., the opening of the handwritten 'A' in **Fig. 9b**). This behavior can potentially be due to the nonlinear interaction of the subwavelength photoconductive microprobe system (**Fig. 7f**) with the local field distribution at the output plane that we experimentally imaged.

Overall, our experiments confirmed the subwavelength imaging capabilities of our solid-immersion diffractive imager, successfully resolving linewidths of $0.293\lambda$ ($\sim\lambda/2n$) while also performing direct (all-optical) phase retrieval of subwavelength features through $P \rightarrow I$ transformations.

**METHODS**

**Forward-propagation model of solid-immersion diffractive imagers:** A solid-immersion diffractive imager consists of a single diffractive encoder layer and $L$ diffractive decoder layers. A dielectric slab of index $n$ is placed between the diffractive encoder layer and the object, which allows the transmission of high-frequency information from the object toward the encoder surface. Forward propagation of the complex electromagnetic field can be modeled as a sequence of (1) free-space propagation between the $l^{th}$ and $(l + 1)^{th}$ diffractive layers (where $l = 0, 1, 2, \ldots, L+1$) including the input plane ($l = 0$), the encoder layer ($l = 1$), the decoder layers ($l = 2, \ldots, L+1$) and the image plane ($l = L+2$), and (2) the modulation of the optical field by the diffractive encoder/decoder layers ($l = 1, \ldots, L+1$). The propagation of the complex optical field in the air and the dielectric medium is modeled by the angular spectrum method [44]. The 2D complex optical field profile $u^l(x,y)$ processed by the $l^{th}$ diffractive layer after propagation over an axial distance of $d$ in a medium with a refractive index $n$ can be calculated by:

$$P_d u^l(x,y) = \mathcal{F}^{-1}\{\mathcal{F}\{u^l(x,y)\}H(f_x, f_y; d; n)\} \tag{1}$$

where the operator $P_d$ represents the free-space propagation, operator $\mathcal{F}$ ($\mathcal{F}^{-1}$) is the two-dimensional (inverse) Fourier transform, and $H(f_x, f_y; d; n)$ is the transfer function of free space ($n = 1$) or the dielectric medium ($n > 1$):

$$H(f_x, f_y; d; n) = \begin{cases} 0, & f_x^2 + f_y^2 > \frac{n^2}{\lambda^2} \\ \exp\left\{jkd\sqrt{1 - \left(\frac{2\pi f_x}{nk}\right)^2 - \left(\frac{2\pi f_y}{nk}\right)^2}\right\}, & f_x^2 + f_y^2 \leq \frac{n^2}{\lambda^2} \end{cases} \tag{2}$$

where $j = \sqrt{-1}$, $\lambda$ is the illumination wavelength, $k = \frac{2\pi}{\lambda}$ is the wavevector, and $f_x$, $f_y$ are the spatial frequencies on the $x$-$y$ plane, orthogonal to the direction of the wave propagation.

We modeled both the diffractive solid-immersion encoder and decoder layers as phase-only modulators of the complex incident fields, where the complex transmittance coefficient $t^l$ of the $l^{th}$ diffractive layer can be written as:

$$t^l(x,y) = \exp\left(j\phi^l(x,y)\right) \tag{3}$$

$\phi^l(x,y)$ represents the phase delay values of the diffractive features on the $l^{th}$ diffractive layer. The 2D complex optical fields at the output/image plane can be derived by combining equations (1) and (3):

$$o(x,y) = P_{d_{L+1,L+2}}\left[\prod_{l=1}^{L+1} t^l(x,y) \cdot P_{d_{l-1,l}}\right] i(x,y) \tag{4}$$

where $d_{l-1,l}$ represents the axial distance between the $(l - 1)^{th}$ and the $l^{th}$ layers, $i(x,y)$ is the input optical field in the $x$-$y$ plane.

**Implementation details and training of solid-immersion diffractive imagers:** In the numerical and experimental demonstrations of this work, the diffractive imagers were trained with a hybrid dataset of (1) grating images as resolution test targets and (2) handwritten letters/digits from the EMNIST (for numerical tests) or the MNIST dataset (for experimental demonstrations). All diffractive encoder/decoder layers had a lateral pixel/neuron size of 0.4 mm and used $\lambda = 0.75$ mm. Unless otherwise stated, the axial distance between the input image/object and the diffractive encoder layer, the distances between successive encoder/decoder layers, and the axial distance between the last diffractive decoder layer and the output plane were set to 0.75 mm ($1\lambda$), 9 mm ($12\lambda$), and 12 mm ($16\lambda$), respectively.

Each diffractive layer contained 120×120 phase-valued diffractive features ($64\lambda \times 64\lambda$) in the *x-y* plane for both numerical and experimental demonstrations. The input sizes of the resolution test targets and other images from various datasets, including MNIST, EMINST, FASHION-MNIST, and QuickDraw were set to 24×24 and 12×12 pixels for solid-immersion diffractive imagers trained for numerical testing and experimental validation, respectively. During the training, each raw image was linearly up-sampled in both *x* and *y* directions by a factor of 2. To model the field propagation process accurately, we also up-sampled the diffractive surfaces by 4 times, i.e., from 120×120 to 480×480 pixels, so that a lateral grid size of 0.1 mm was consistently used for all the calculations in the forward model. Lastly, the input images and the diffractive encoder/decoder layers were zero-padded to 520×520 pixels (in the *x-y* plane) for forward model calculations.

The resolution test target image datasets were created as images of straight and curved lines, where the height profile follows a sine function ranging from 0 to 1. For the training and validation datasets, we used 8000 and 2000 images of lines with randomly defined curvature and linewidths ($w = 0.15$-$0.4$ mm), respectively. For blind testing, the test image dataset size corresponding to each linewidth/resolution was selected as 100 images. The images of EMNIST (for numerical tests) or MNIST (for experimental tests) datasets were each divided into training, validation, and testing datasets without overlap, with each set containing 48,000, 12,000, and 10,000 images, respectively. During the training process, hybrid datasets combining the resolution test targets with the EMNIST (for numerical tests) or MNIST (for experimental demonstrations) datasets were created by merging the corresponding training and validation datasets.

The diffractive models were optimized *via* a stochastic gradient-based error back-propagation process using the Adam optimizer [47] to minimize the user-defined loss function, Eq. (8), with a learning rate of 0.002. The batch size was selected as 30. The diffractive models were trained and tested using PyTorch 1.12 or 1.13 with a single GeForce RTX 3080/3090 graphical processing unit (GPU, from Nvidia Inc.). The typical training time of a solid-immersion diffractive imager for 1000 epochs is ~8 hours.

**Fabrication of monolithic solid-immersion diffractive imagers:** The transmissive layer designs of subwavelength diffractive imagers were converted into an STL file with MATLAB. The monolithic design of this multi-layered structure was obtained in a single printing session by a 3D printer (Objet30 Pro, Stratasys Ltd.) using the ultraviolet curable material (VeroBlackPlus RGD875, Stratasys Ltd.) as the ink/printing material and the PolyJet's gel-like support material (SUP705B, Stratasys Ltd) as the support between the diffractive layers. The support materials were removed with a combination of mechanical rubbing and chemical washing (see Supplementary Fig. S10 for details). First, we mechanically removed the majority of the support materials between the layers with a metallic stick and flushed the layers with waterjet for 3 minutes. Then, the sample was placed in 500 mL of 5% KOH solution for 5 hours to soften the residual support materials and then rinsed with excess DI water for 10 minutes to fully remove the KOH. Lastly, the softened residual support materials were removed by another waterjet, and the cleaned diffractive imager was air-dried.

**Supplementary Information** includes:

- Training Loss Functions

- Experimental set-up

Corresponding Author: E-mail: ozcan@ucla.edu. (A. Ozcan)  ORCID: 0000-0002-0717-683X

# REFERENCES

1. P. Chakravarthula, et al., Learned hardware-in-the-loop phase retrieval for holographic near-eye displays. J ACM Trans. Graph., **39**(6 %), 186 (2020)
2. Y. Jo, et al., Quantitative phase imaging and artificial intelligence: A review. IEEE Journal of Selected Topics in Quantum Electronics, **25**(1), 1-14 (2019)
3. H. Majeed, et al., Quantitative phase imaging for medical diagnosis. Journal of Biophotonics, **10**(2), 177-205 (2017)
4. Y. Park, C. Depeursinge, G. Popescu, Quantitative phase imaging in biomedicine. Nature Photonics, **12**(10), 578-589 (2018)
5. U. S. Kamilov, et al., Learning approach to optical tomography. Optica, **2**(6), 517-522 (2015)
6. M. A. Beltran, D. M. Paganin, K. Uesugi, M. J. Kitchen, 2D and 3D x-ray phase retrieval of multi-material objects using a single defocus distance. Opt. Express, **18**(7), 6423-6436 (2010)
7. A. Descloux, et al., Combined multi-plane phase retrieval and super-resolution optical fluctuation imaging for 4D cell microscopy. Nature Photonics, **12**(3), 165-172 (2018)
8. M. Wan, J. J. Healy, J. T. Sheridan, Terahertz phase imaging and biomedical applications. Optics & Laser Technology, **122**105859 (2020)
9. F. Zernike, How i discovered phase contrast. Science, **121**(3141), 345-349 (1955)
10. W. Lang, *Nomarski differential interference-contrast microscopy*. Carl Zeiss Oberkochen: 1982.
11. M. K. Sharma, et al., Inverse scattering via transmission matrices: Broadband illumination and fast phase retrieval algorithms. IEEE Transactions on Computational Imaging, **6**95-108 (2020)
12. R. K. Singh, A. M. Sharma, B. Das, Quantitative phase-contrast imaging through a scattering media. Opt. Lett., **39**(17), 5054-5057 (2014)
13. T. Wu, J. Dong, S. Gigan, Non-invasive single-shot recovery of a point-spread function of a memory effect based scattering imaging system. Opt. Lett., **45**(19), 5397-5400 (2020)
14. S. Yoon, et al., Deep optical imaging within complex scattering media. Nature Reviews Physics, **2**(3), 141-158 (2020)
15. C. W. Mccutchen, Superresolution in microscopy and the abbe resolution limit. J. Opt. Soc. Am., **57**(10), 1190-1192 (1967)
16. J. Gass, A. Dakoff, M. K. Kim, Phase imaging without $2\pi$ ambiguity by multiwavelength digital holography. Opt. Lett., **28**(13), 1141-1143 (2003)
17. C. J. Mann, P. R. Bingham, V. C. Paquit, K. W. Tobin, Quantitative phase imaging by three-wavelength digital holography. Opt. Express, **16**(13), 9753-9764 (2008)
18. C. J. Mann, L. Yu, C.-M. Lo, M. K. Kim, High-resolution quantitative phase-contrast microscopy by digital holography. Opt. Express, **13**(22), 8693-8698 (2005)
19. J. Park, et al., Artificial intelligence-enabled quantitative phase imaging methods for life sciences. Nature Methods, **20**(11), 1645-1660 (2023)
20. B. Javidi, et al., Roadmap on digital holography. Opt. Express, **29**(22), 35078-35118 (2021)
21. Y. Rivenson, Y. Wu, A. Ozcan, Deep learning in holography and coherent imaging. Light: Science & Applications, **8**(1), 85 (2019)
22. G. Zheng, R. Horstmeyer, C. Yang, Wide-field, high-resolution fourier ptychographic microscopy. Nature Photonics, **7**(9), 739-745 (2013)
23. L. Tian, L. Waller, 3D intensity and phase imaging from light field measurements in an led array microscope. Optica, **2**(2), 104-111 (2015)

24. N. V. Chernomyrdin, et al., Reflection-mode continuous-wave 0.15λ-resolution terahertz solid immersion microscopy of soft biological tissues. Applied Physics Letters, **113**(11), (2018)
25. A. Darafsheh, et al., Advantages of microsphere-assisted super-resolution imaging technique over solid immersion lens and confocal microscopies. Applied Physics Letters, **104**(6), (2014)
26. L. Sapienza, M. Davanço, A. Badolato, K. Srinivasan, Nanoscale optical positioning of single quantum dots for bright and pure single-photon emission. Nature Communications, **6**(1), 7833 (2015)
27. M. Totzeck, W. Ulrich, A. Göhnermeier, W. Kaiser, Pushing deep ultraviolet lithography to its limits. Nature Photonics, **1**(11), 629-631 (2007)
28. S. M. Mansfield, G. S. Kino, Solid immersion microscope. Applied Physics Letters, **57**(24), 2615-2616 (1990)
29. B. D. Terris, et al., Near‐field optical data storage using a solid immersion lens. Applied Physics Letters, **65**(4), 388-390 (1994)
30. Q. Wu, G. D. Feke, R. D. Grober, L. P. Ghislain, Realization of numerical aperture 2.0 using a gallium phosphide solid immersion lens. Applied Physics Letters, **75**(26), 4064-4066 (1999)
31. G. J. Brakenhoff, P. Blom, P. Barends, Confocal scanning light microscopy with high aperture immersion lenses. Journal of Microscopy, **117**(2), 219-232 (1979)
32. W. T. Chen, et al., Immersion meta-lenses at visible wavelengths for nanoscale imaging. Nano Letters, **17**(5), 3188-3194 (2017)
33. P. Marquet, et al., Digital holographic microscopy: A noninvasive contrast imaging technique allowing quantitative visualization of living cells with subwavelength axial accuracy. Opt. Lett., **30**(5), 468-470 (2005)
34. S. Chowdhury, W. J. Eldridge, A. Wax, J. A. Izatt, Structured illumination multimodal 3d-resolved quantitative phase and fluorescence sub-diffraction microscopy. Biomed Opt Express, **8**(5), 2496-2518 (2017)
35. C. Zuo, et al., High-resolution transport-of-intensity quantitative phase microscopy with annular illumination. Scientific Reports, **7**(1), 7654 (2017)
36. C. Zheng, et al., High spatial and temporal resolution synthetic aperture phase microscopy. Advanced Photonics, **2**(6), 065002 (2020)
37. H. Xiao, K. Rasul, R. Vollgraf, Fashion-mnist: A novel image dataset for benchmarking machine learning algorithms. arXiv preprint **arXiv:1708.07747** (2017)
38. J. Jongejan, et al., The quick, draw!-ai experiment. Mount View, CA, accessed Feb, **17**(2018), 4 (2016)
39. X. Lin, et al., All-optical machine learning using diffractive deep neural networks. Science, **361**(6406), 1004-1008 (2018)
40. D. Mengu, Y. Luo, Y. Rivenson, A. Ozcan, Analysis of diffractive optical neural networks and their integration with electronic neural networks. IEEE Journal of Selected Topics in Quantum Electronics, **26**(1), 1-14 (2019)
41. E. Kussul, T. Baidyk, Improved method of handwritten digit recognition tested on mnist database. Image and Vision Computing, **22**(12), 971-981 (2004)
42. D. Mengu, et al., Misalignment resilient diffractive optical networks. Nanophotonics, **9**(13), 4207-4219 (2020)

43. N. T. Yardimci, S. H. Yang, C. W. Berry, M. Jarrahi, High-power terahertz generation using large-area plasmonic photoconductive emitters. IEEE Transactions on Terahertz Science and Technology, **5**(2), 223-229 (2015)
44. E. Wolf, Electromagnetic diffraction in optical systems-i. An integral representation of the image field. Proceedings of the Royal Society of London. Series A. Mathematical, **253**(1274), 349-357 (1959)
45. D. Mengu, et al., At the intersection of optics and deep learning: Statistical inference, computing, and inverse design. Adv. Opt. Photon., **14**(2), 209-290 (2022)
46. N. U. Dinc, et al., Computer generated optical volume elements by additive manufacturing. **9**(13), 4173-4181 (2020)
47. D. P. Kingma, J. Ba, Adam: A method for stochastic optimization. arXiv preprint, **arXiv: 1412.6980** (2014)

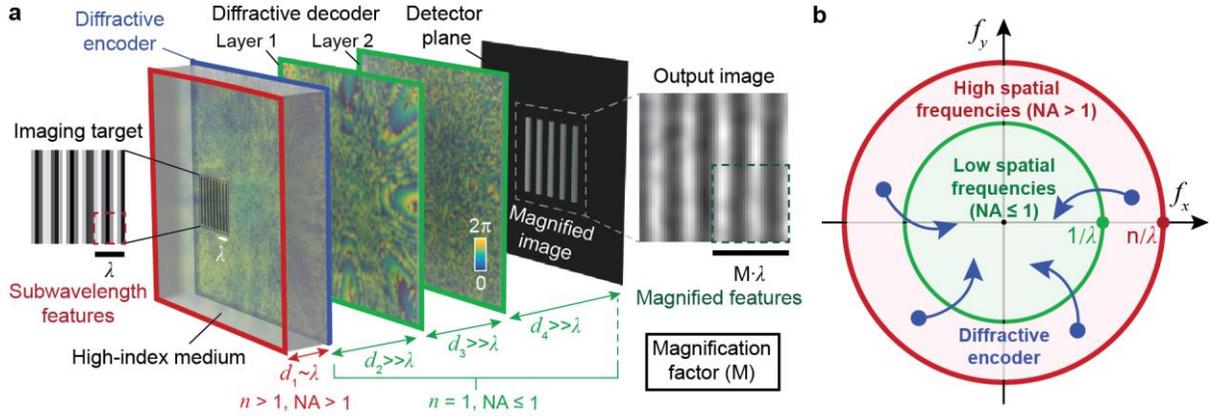

**Figure 1: Subwavelength imaging using a solid-immersion diffractive optical processor.** (**a**) Scheme showing the design of the subwavelength imager consisting of a diffractive solid-immersion encoder and successive decoder layers that axially span $<100\lambda$ between the sample and the output image plane. (**b**) Frequency-domain diagram illustrating the transformation of high frequency information ($f > 1/\lambda$, which can only propagate in a high-index medium, n > 1) towards lower frequency points ($f \lesssim 1/\lambda$, which can propagate in air). The high-index solid-immersion medium ($n > 1$) in (a) is between the object and the diffractive encoder. The diffractive decoder reconstructs an output image, magnified compared to the original image by a factor of M.

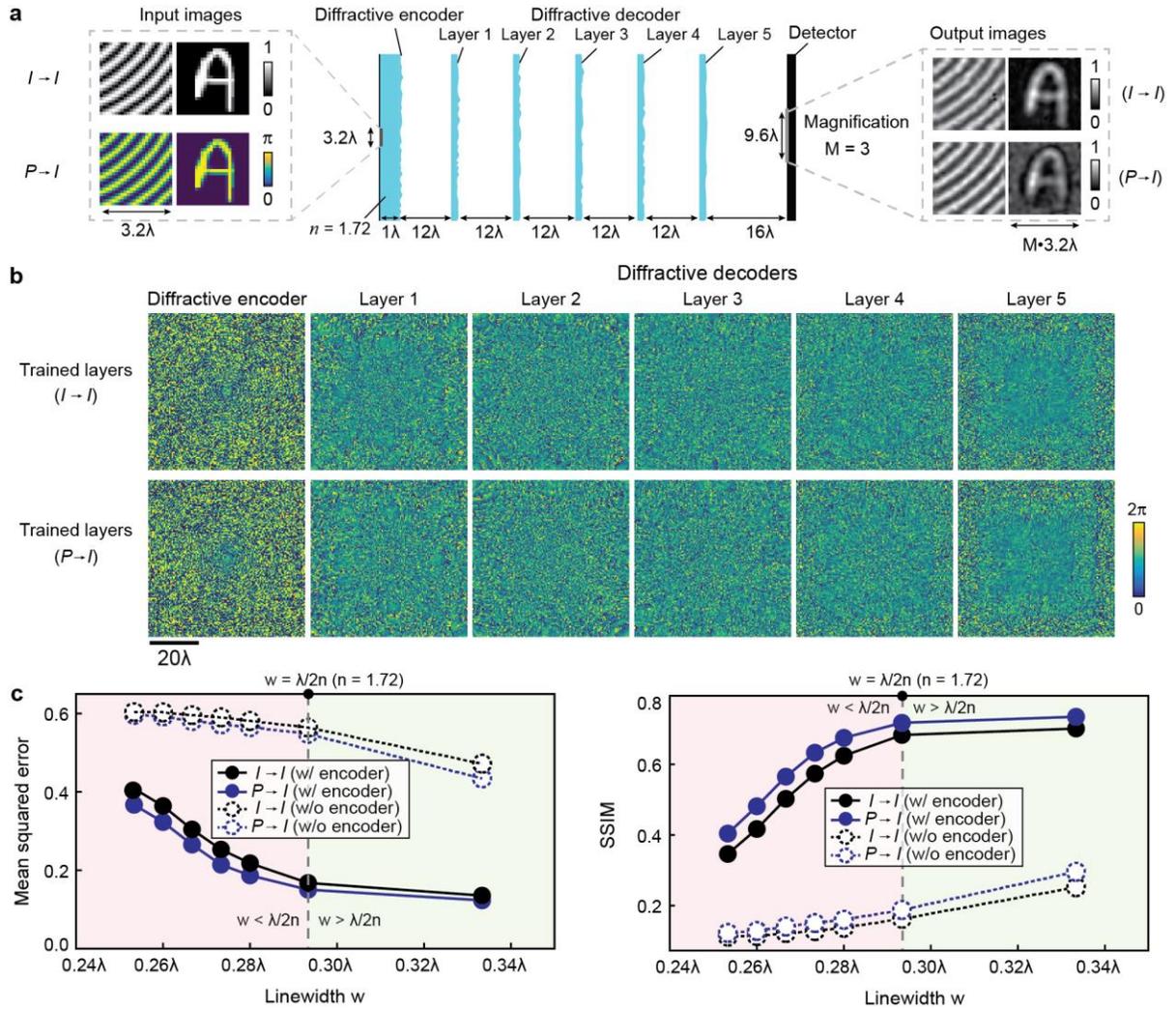

**Figure 2**: **Design and performance analysis of solid-immersion diffractive optical processors for intensity- and phase-encoded input objects.** (**a**) Scheme depicting a diffractive imager with a magnification factor of M = 3 using $L = 5$ diffractive decoder layers. (**b**) Phase profiles of the diffractive encoder and decoder layers optimized *via* a deep-learning training process for (upper) intensity- and (lower) phase-encoded objects, respectively. (**c**) Performance analysis of solid-immersion diffractive optical imagers for intensity- and phase-encoded resolution test targets with different linewidths ranging from ~0.253$\lambda$ to ~0.333$\lambda$ using mean squared error (MSE) and structural similarity index measure (SSIM). The results in (c) are also compared to baseline designs, where the decoder layers were trained without an encoder (also see Supplementary Fig. S1).

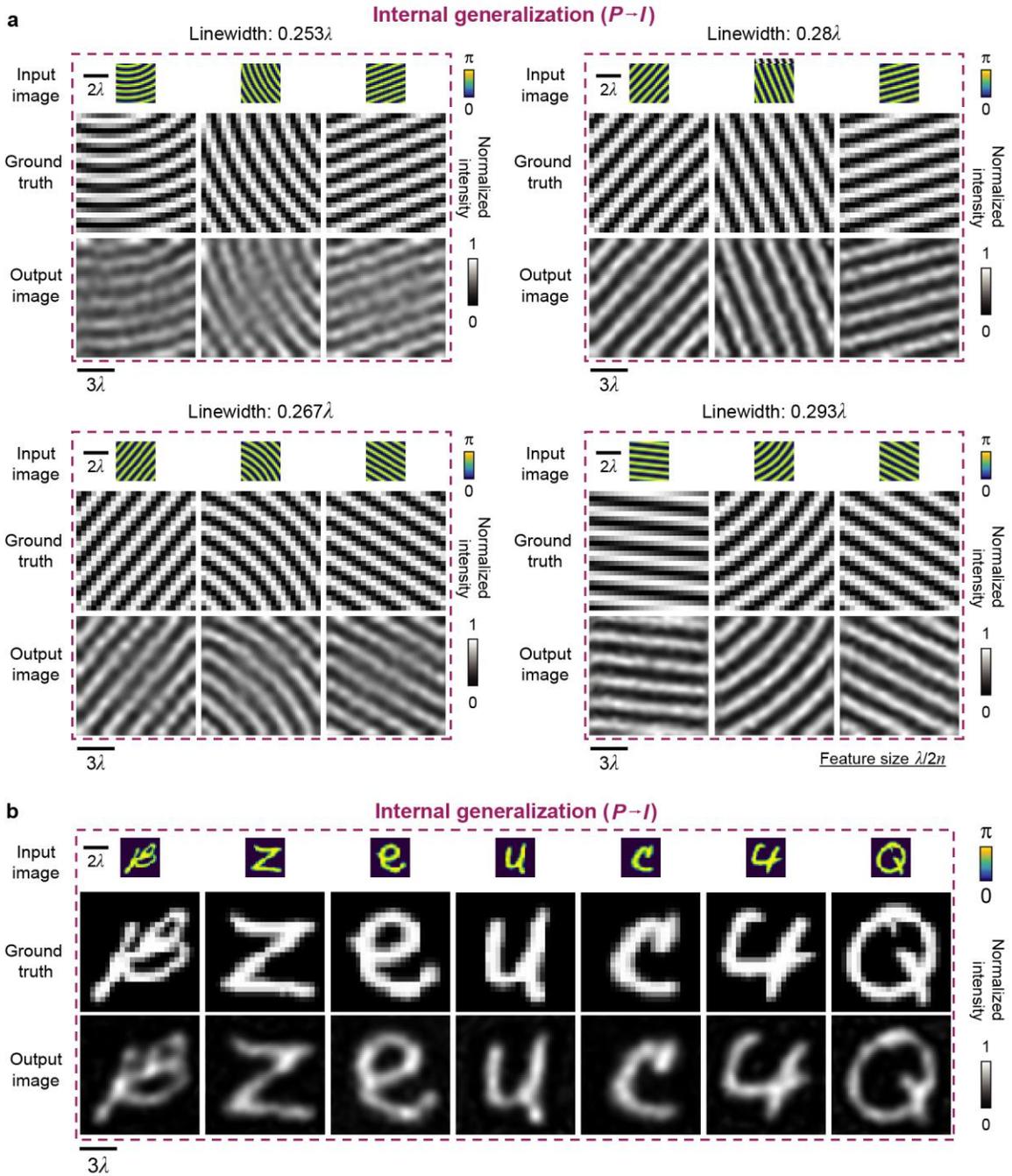

**Figure 3**: **Internal generalization of the solid-immersion diffractive imager to unknown phase-encoded objects ($P \rightarrow I$ transformations)**. (**a**) Imaging results using (**a**) resolution-test targets with linewidths ranging from ~0.253λ to ~0.293λ and (**b**) EMNIST handwritten letters/digits. The diffractive processor consists of 1 encoder layer and $L = 5$ decoder layers, all jointly optimized (see **Fig. 2**).

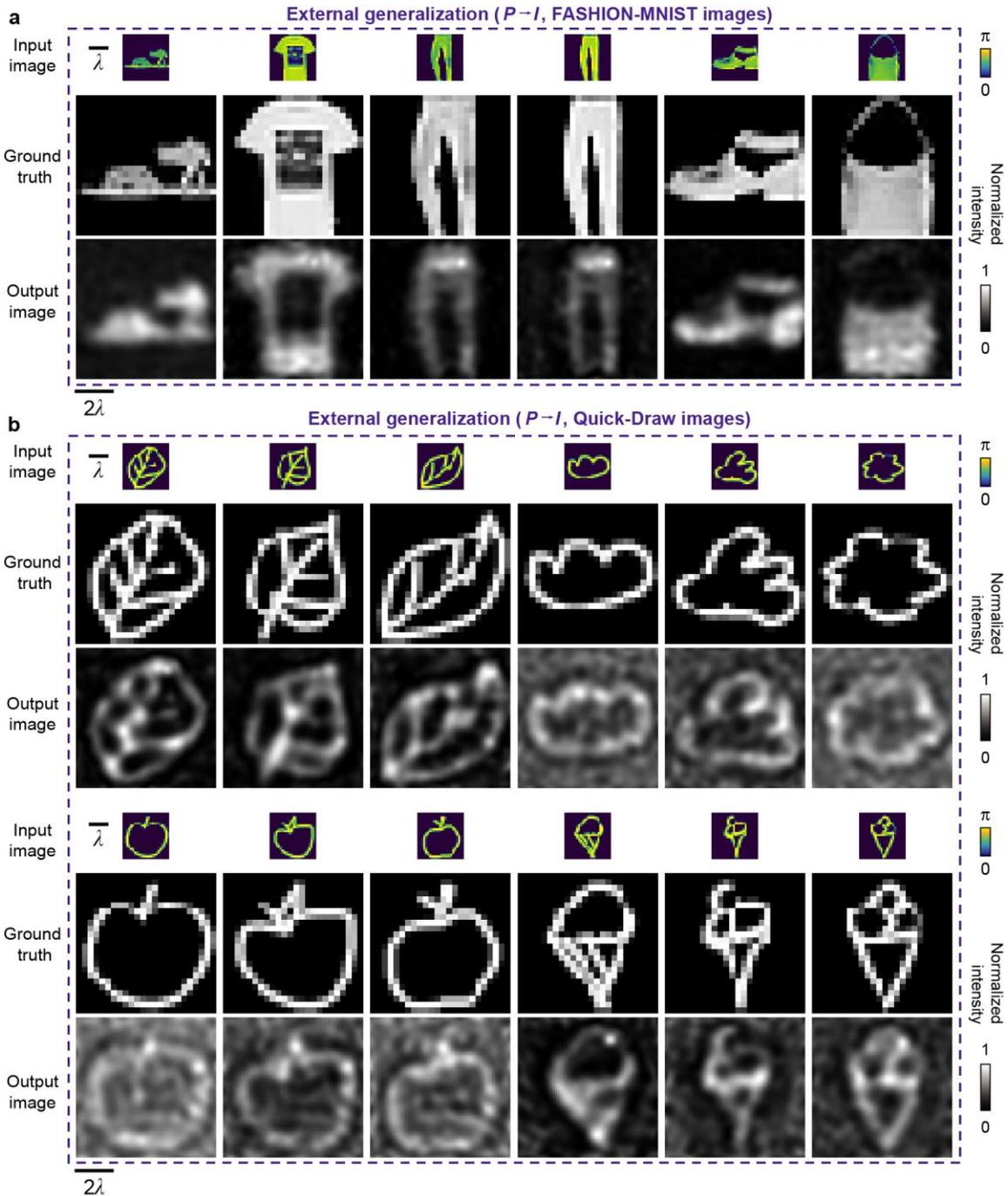

**Figure 4**: **External generalization of the solid-immersion diffractive imager to new types of objects from unknown image datasets for $P \rightarrow I$ transformations.** Blind testing results using (**a**) Fashion-MNIST and (**b**) Quick-Draw image datasets. The diffractive processor consists of 1 encoder layer and $L$=5 decoder layers, all jointly optimized (see **Fig. 2**). The diffractive model was trained with various gratings and EMNIST images.

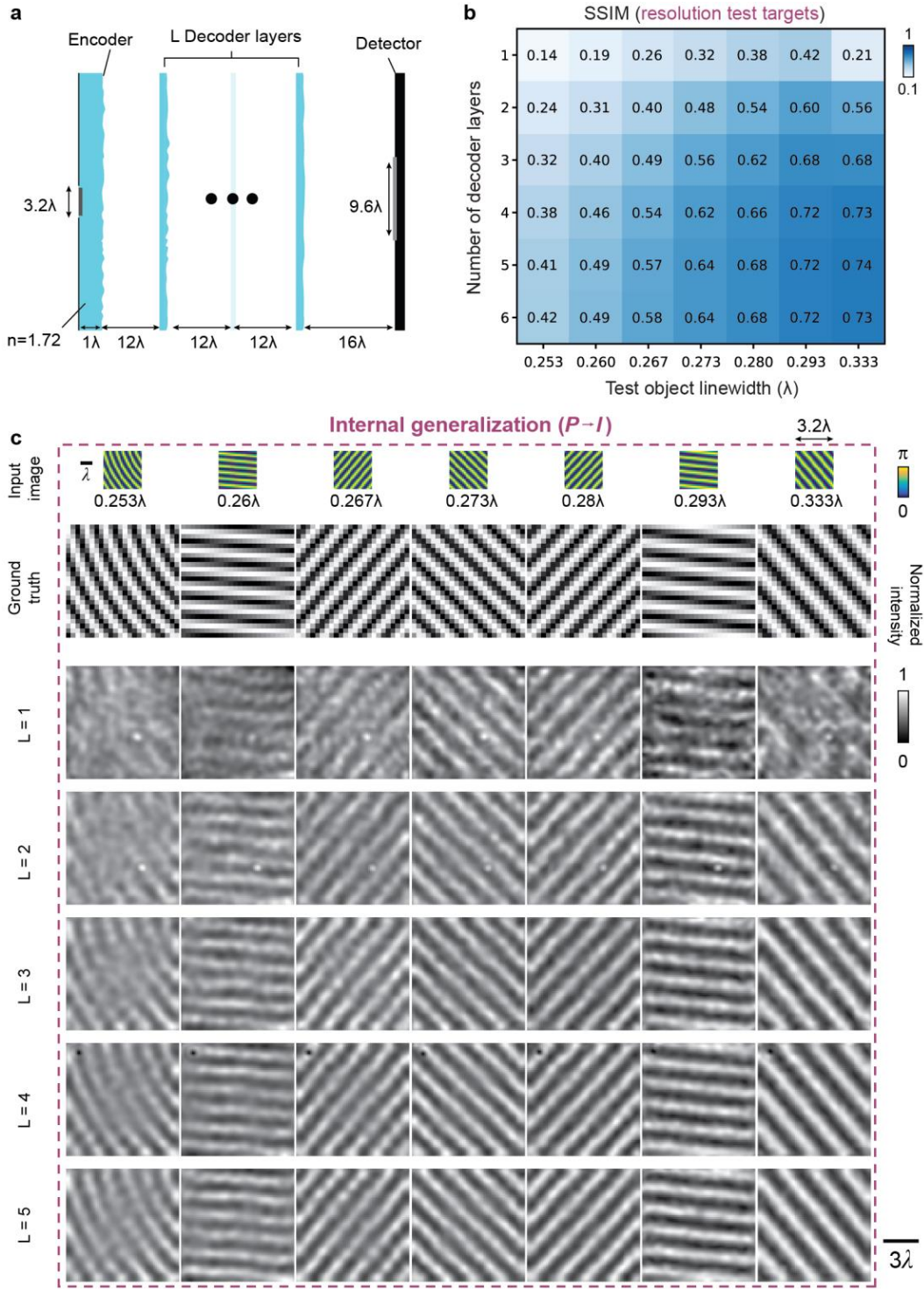

**Figure 5**: **Resolution tests for solid-immersion diffractive imagers designed with different numbers of decoder layers.** (**a**) Scheme depicting the general design of the diffractive subwavelength imager network with an arbitrary number (*L*) of diffractive decoder layers. (**b**) Quantitative analysis and (**c**) output images showing the all-optical reconstructions for resolution test targets achieved by diffractive imagers with $L = 1-5$ decoder layers.

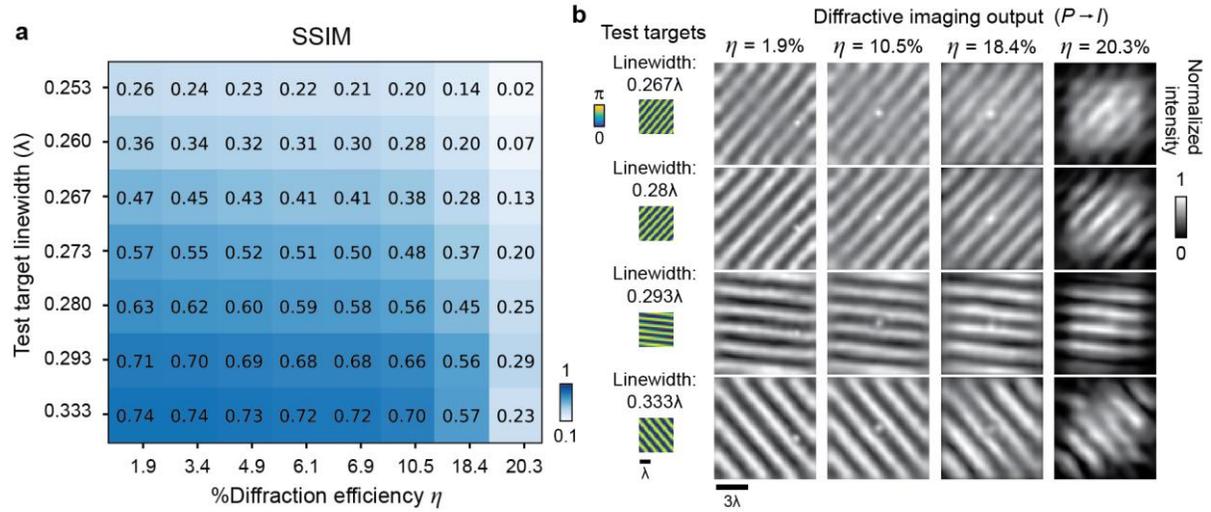

**Figure 6**: **Trade-off between the output diffraction efficiency ($\eta$) and the imaging performance**. (**a**) Calculated SSIM values and (**b**) selected blind testing results evaluated using phase-encoded resolution test targets with linewidths ranging from ~0.253$\lambda$ to ~0.333$\lambda$ for diffractive imagers trained to exhibit different output diffraction efficiencies. The diffractive processor consisted of 1 encoder layer and 5 decoder layers. The diffractive model was trained with images of phase gratings (linewidths ~0.2-0.53$\lambda$) and EMNIST images that excluded the test images used in (a-b).

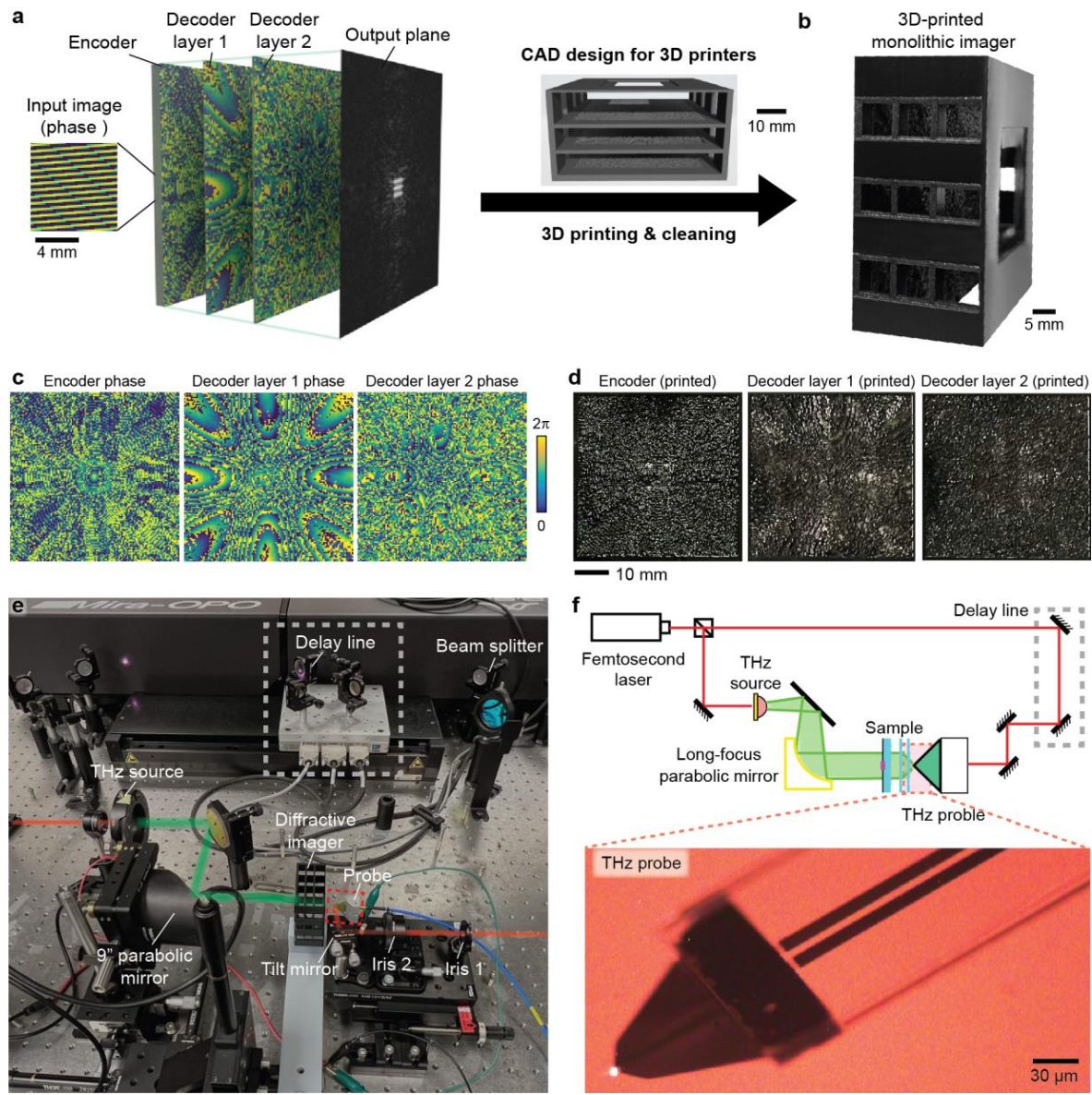

**Figure 7**: **Fabrication method for solid-immersion diffractive imagers and the experimental setup.** (**a**) Design of the multi-layered subwavelength imager consisting of a diffractive encoder and two decoder layers and their CAD design. (**b**) The 3D-printed monolithic imager after the cleaning process. (**c**) Trained phase profiles of the encoder and decoder layers of the subwavelength diffractive imager. (**d**) Fabricated layers of the diffractive encoder and decoder layers. (**e**) Photograph of the THz-TDS experimental setup. (**f**) Top: schematic of the THz-TDS setup. Red lines represent the optical path of the femtosecond pulses (central wavelength: 800 nm). Green lines represent the optical path of the terahertz wave (peak frequency, ~500 GHz, observable bandwidth, ~5 THz). Bottom: an optical image of the THz microprobe with a photoconductive gap size (at tip of the microprobe) of 2 μm.

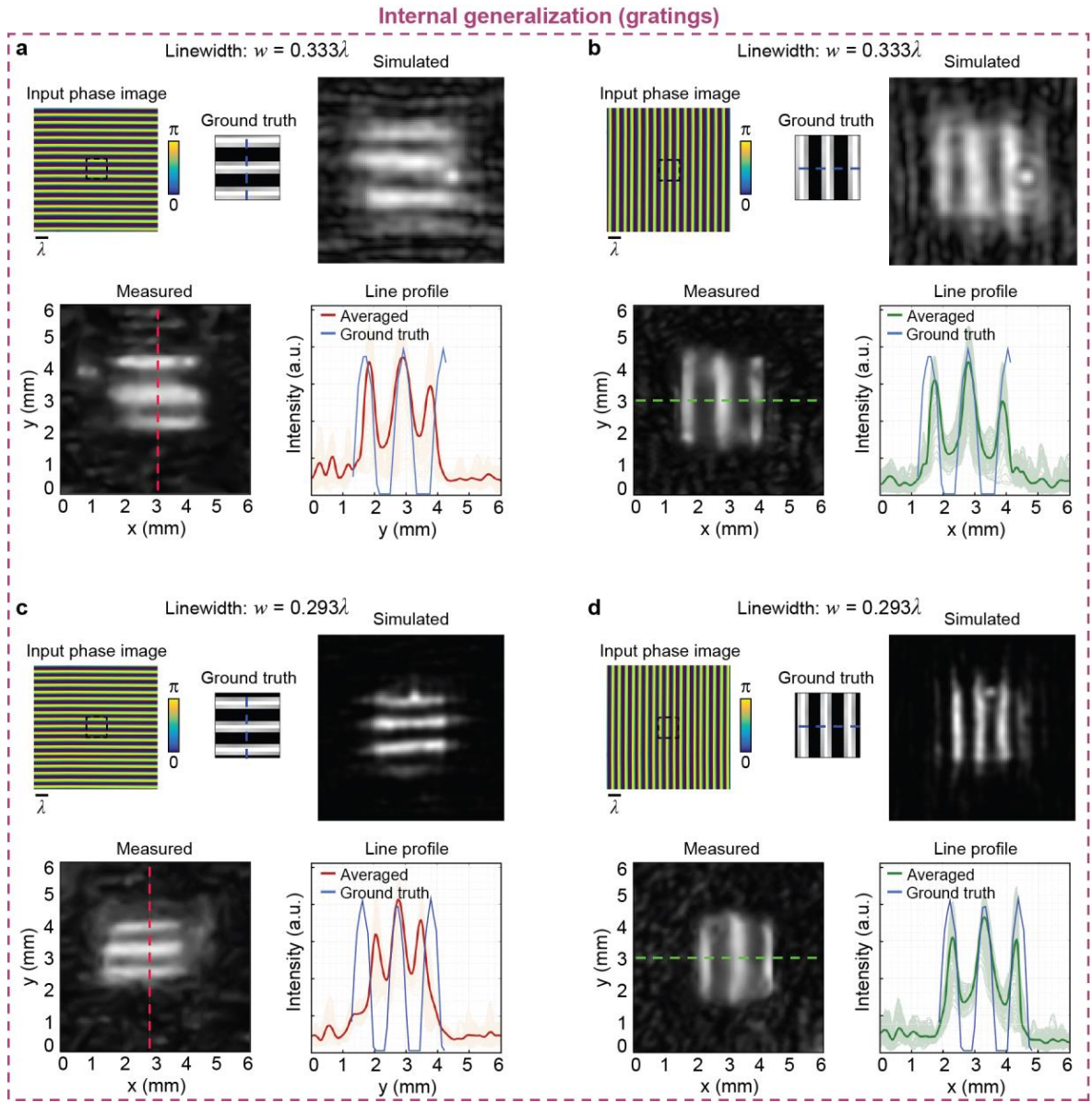

**Figure 8: Experimental demonstrations of ($P \rightarrow I$) transformations corresponding to phase-encoded resolution test targets with subwavelength resolution.** (Upper left) Input phase images, (upper middle) ground truth, (upper right) simulated imaging results, (lower left) measured intensity images at the output plane, and (lower right) the corresponding line profiles of the resolution-test targets with a linewidth of ~0.333$\lambda$ oriented in (**a**) *x* and (**b**) *y* directions as well as the resolution-test targets with a linewidth of ~0.293$\lambda$ oriented in (**c**) *x* and (**d**) *y* directions.

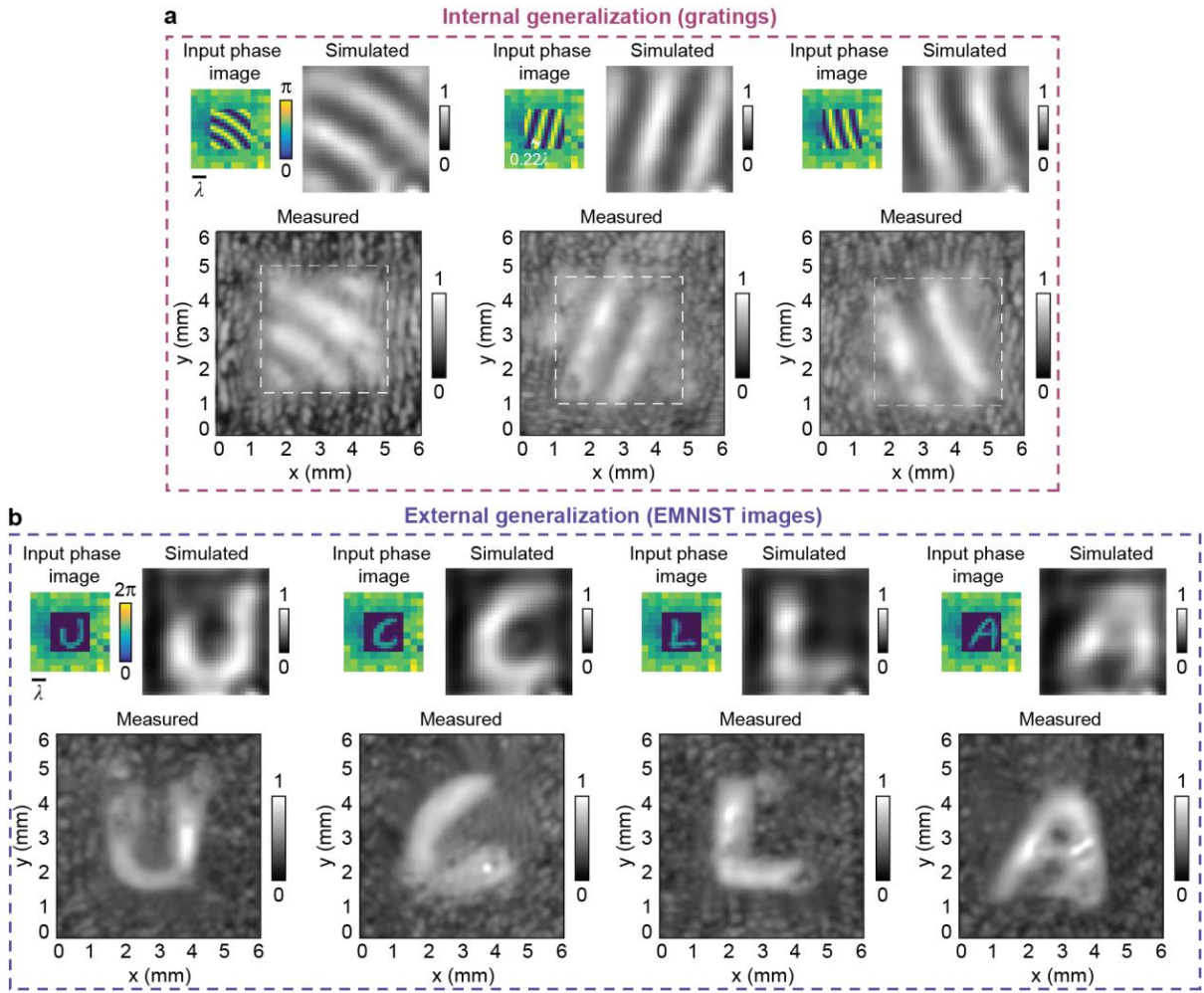

**Figure 9**: **Experimental demonstrations of ($P \rightarrow I$) transformations corresponding to phase-encoded gratings and EMNIST handwritten letters with subwavelength resolution.** (**a**) Imaging results of different gratings with a linewidth of ~0.293$\lambda$; (**b**) imaging results of EMNIST handwritten letters "U", "C", "L", "A", demonstrating the ability of the diffractive processor to image new types of objects never seen before (external generalization). (Upper left) Input phase images, (upper right) simulated intensity images, and (lower) the experimentally measured output images.